\documentclass[pra,onecolumn,superscriptaddress,floatfix]{revtex4}

\usepackage{amsmath}
\usepackage{times}
\usepackage{graphicx}

\mathchardef\ogon="012C%
\newcommand{\as}{a\kern-0.22em\lower.40ex\hbox{$_{\ogon}$}}

\begin{document}

\title{Dynamics of correlations in atomic Bose-Einstein condensates}

\author{Krzysztof G{\'o}ral}
\affiliation{Clarendon Laboratory, Department of Physics, 
University of Oxford, Parks Road, Oxford, OX1 3PU, United Kingdom}
\affiliation{Center for Theoretical Physics, Polish Academy of 
Sciences, Aleja Lotnik\'ow 32/46, 02-668 Warsaw, Poland}
\author{Thorsten K\"{o}hler}
\affiliation{Clarendon Laboratory, Department of Physics, 
University of Oxford, Parks Road, Oxford, OX1 3PU, United Kingdom}
\author{Thomas Gasenzer}
\affiliation{Institut f\"{u}r Theoretische Physik, Universit\"{a}t
  Heidelberg, Philosophenweg 16, D-69120 Heidelberg, Germany}
\author{Keith Burnett}
\affiliation{Clarendon Laboratory, Department of Physics, 
University of Oxford, Parks Road, Oxford, OX1 3PU, United Kingdom}

\begin{abstract}
The Gross-Pitaevskii equation has been extremely successful in the
theory of weakly-interacting Bose-Einstein condensates. However,
present-day experiments reach beyond the regime of its validity due to
the significant role of correlations. We review a method of tackling the
dynamics of correlations in Bose condensed gases, in terms of
noncommutative cumulants. This new approach has a wide range
of applicability in the areas of current interest, e.g. the production of
molecules and the manipulation of interactions in condensates. It also
offers an interesting perspective on the classical-field methods for partly
condensed Bose gases.
\end{abstract}

\maketitle

Eight years after the first experimental achievement of Bose-Einstein
condensation of atoms \cite{Anderson95}, the field of quantum-degenerate
gases is not losing its pace. On the contrary, it continues to open
new and promising avenues for research, with the latest hot topics
including quantum vortices \cite{Matthews99,Madison00}, Bose condensation
of molecules \cite{Jochim03} and prospects of elucidating pairing
mechanisms with the help of ultracold Fermi gases \cite{Levi03}.

Given the enormous potential and diversity of the field, it comes as
a surprise that most of the experiments performed so far with
Bose-Einstein condensates have been modelled almost perfectly by
mean-field theory. The famous Gross-Pitaevskii equation
\cite{Pitaevskii61,Gross61} has proved to be a genuine workhorse in the area, being
employed in a myriad of contexts \cite{Dalfovo99}. An interesting and
relatively new addition to the plethora of applications is a classical-field
method for the dynamics of interacting Bose gases at non-zero
temperatures
\cite{Svistunov91,Damle96,Goral01,Davis01,Sinatra01,Goral02,Schmidt03}.

Only recently have new types of experiments been performed which
necessitate the use of more advanced theoretical techniques. A
striking example is provided by an experimental demonstration of a
quantum phase transition from a superfluid to a Mott insulator with
bosons in an optical lattice \cite{Greiner02}. In this case, the mean field
vanishes completely in the insulating phase. Mean-field theory also
fails in a strongly-interacting regime accessible through
magnetically tunable two-body scattering in the vicinity of a Feshbach resonance
\cite{Inouye98,Donley02}. Very recently, a number of experiments have
explored the association of ultracold molecules through crossing a Feshbach
resonance \cite{Herbig03,Xu03,Duerr03}. Again, in this situation mean field
approaches provide an approximate description of the processes
involved in limiting cases only \cite{KGTKSAGETPSJ03}.

Apart from the availability of new experimental results challenging
the applicability of the lowest order, mean-field techniques, a very
interesting general issue is to what extent the time-dependent
Gross-Pitaevskii equation describes a dynamical response of a
condensed gas to perturbations. Experiments operating in the vicinity
of  a Feshbach resonance provide just such an example where
external perturbations are violent and may lead to an almost
complete depletion of the condensate. However, one can imagine other,
more general situations, e.g. a rapid variation of a trapping
potential, where losses can also be expected to be significant.

In this paper, we review a recently developed method capable of
handling the dynamics of a Bose condensed gas beyond mean field theory
in a self-consistent, number-conserving way \cite{TKKB02}. The method is based on the
expansion of correlation functions in terms of noncommutative
cumulants \cite{Fricke96}. After sketching the general ideas behind
this new approach, we describe a test of its predictions against the
experimental data in a particular case of a Feshbach resonance
crossing performed with a $^{85}$Rb condensate at JILA
\cite{Cornish00}.
Furthermore, we provide an outlook for other possible applications of the cumulant approach.

We start from a very general form of a many-body Hamiltonian in its
second-quantised form:

  \begin{equation}
    \label{Hamiltonian}
  H=\int d\mathbf{x} \ \psi^\dagger(\mathbf{x})
  H_\mathrm{1B}(\mathbf{x})\psi(\mathbf{x})
  +\frac{1}{2}
  \int d\mathbf{x}d\mathbf{y} \ \psi^\dagger(\mathbf{x})
  \psi^\dagger(\mathbf{y})V(\mathbf{x}-\mathbf{y})
  \psi(\mathbf{x})\psi(\mathbf{y}) \, ,
\end{equation}

\noindent where $\psi(\mathbf{x})$ is a field operator for atoms, obeying the
bosonic commutation relations:
$[\psi(\mathbf{x}),\psi^\dagger(\mathbf{y})]=\delta(\mathbf{x}-\mathbf{y})$
and
$[\psi(\mathbf{x}),\psi(\mathbf{y})]=0$. $V(\mathbf{x}-\mathbf{y})$ is
the two-body interaction potential. The one-body Hamiltonian, which may
contain an external (e.g. trapping) potential, is defined as follows:

\begin{equation}
H_\mathrm{1B}=-\hbar^2\nabla^2/2m+V_\mathrm{ext} \, .
\end{equation}

All physical properties of a many-body system, e.g. a gas, are
determined by correlation functions, i.e.  expectation values of
normal ordered products of field operators in the state at time $t$
(denoted hereafter as $\langle \hdots \rangle_t$) \cite{note}. A
usual way of proceeding with the problem of the time evolution of
correlations consists in writing the hierarchy of the equations
of motion, coupled due to the quartic interaction term in the
Hamiltonian. Here we describe a different procedure, which involves
noncommutative cumulants. Given a set of bosonic field operators
$B_1$, $B_2$, $B_3$, ... the cumulants (denoted as $\langle \hdots
\rangle^{c}$) are defined recursively in the following way:

\begin{eqnarray}
\langle B_{1}\rangle_t &=&\langle B_{1}\rangle ^{c},  \nonumber \\
\langle B_{1}B_{2}\rangle_t &=&\langle B_{1}B_{2}\rangle ^{c}+\langle
B_{1}\rangle ^{c}\langle B_{2}\rangle ^{c},  \nonumber \\
\langle B_{1}B_{2}B_{3}\rangle_t &=&\langle B_{1}B_{2}B_{3}\rangle
^{c}+\langle B_{1}\rangle ^{c}\langle B_{2}B_{3}\rangle ^{c} + \langle B_{2}\rangle ^{c}\langle B_{1}B_{3}\rangle ^{c}+\langle
B_{3}\rangle ^{c}\langle B_{1}B_{2}\rangle ^{c}+\langle B_{1}\rangle
^{c}\langle B_{2}\rangle ^{c}\langle B_{3}\rangle ^{c},  \nonumber \\
& \vdots
\label{defcumI}
\end{eqnarray}

It follows from the above definition that the cumulants capture the
{\it essential} correlations in the system by subtracting
factorisable contributions from the correlation functions. They decrease with an
increasing order, provided that the system is not too far away from the
interaction-free equilibrium, which is typically not true for 
correlation functions. In particular, for an ideal gas in
thermal equilibrium all cumulants containing more than two field
operators vanish, in agreement with Wick's theorem of statistical
mechanics. Thus, higher-order cumulants measure the
deviation of the system from the interaction-free situation.

Specifying the cumulant expansion in the case of a partly condensed
gas of bosons, one expects a special role to be played by the
following three lowest-order cumulants:

\begin{eqnarray}
\Psi(\mathbf{x},t)&=&\langle \psi(\mathbf{x}) \rangle_t \, ,\\
\Phi(\mathbf{x},\mathbf{y},t)&=&\langle\psi(\mathbf{y}) \psi(\mathbf{x}) \rangle_t - 
\Psi(\mathbf{y},t) \Psi(\mathbf{x},t) \, ,\\
\Gamma(\mathbf{x},\mathbf{y},t)&=&\langle\psi^{\dagger}(\mathbf{y}) \psi(\mathbf{x}) \rangle_t - 
\Psi^*(\mathbf{y},t)\Psi(\mathbf{x},t) \, .
\end{eqnarray}

The first-order cumulant $\Psi(\mathbf{x},t)$ is the mean field, while
the second order cumulants $\Phi(\mathbf{x},\mathbf{y},t)$ and
$\Gamma(\mathbf{x},\mathbf{y},t)$ will be referred to as the pair
function and the density matrix of the non-condensed fraction,
respectively. We note that the commonly used Gross-Pitaevskii
formalism with the contact potential would involve factorising all correlation functions, which
corresponds to neglecting all cumulants of order higher than one.

The next step is to write the equations of motion for the
cumulants. In order to handle the resulting infinite hierarchy of
them, we define the following truncation scheme of order $n$. The
equations of motion for cumulants containing not more
than $n$ operators are kept exactly. In the equations of motion for
the cumulants of order $n+1$ and $n+2$ we include free evolution only,
i.e. we neglect those products of normal-ordered cumulants that
contain $n+3$ and $n+4$ field operators. 
As a result, the equations of motion
for the cumulants of order $n+1$ and $n+2$ can be solved formally and
the solutions can be substituted back to the equations of motion for
cumulants of order not greater than $n$. The consequence of such a
procedure is that bare microscopic potentials are connected to
transition matrices in all equations of motion up to order $n$.

In what follows, we present the first-order ($n=1$) cumulant approach for
bosons \cite{TKKB02}. In this case the relevant equations of motion, after the
first-order truncation, take the following form:

\begin{eqnarray}
i\hbar \frac{\partial }{\partial t}\Psi ({\bf x},t)&=&H_{{\rm 1B}}({\bf x}
)\Psi ({\bf x},t)+\int d{\bf y}V({\bf x}-{\bf y})[{\langle \psi ^{\dagger }({\bf y})\psi (
{\bf y})\psi ({\bf x})\rangle^{c}}+\Psi ({\bf y},t){ \Gamma ({\bf x},{\bf y
},t)}+\Psi ({\bf x},t){ \Gamma ({\bf y},{\bf y},t)}]  \nonumber \\
&+&\int d{\bf y}V({\bf x}-{\bf y})\Psi ^{\ast }({\bf y},t)\left[ \Phi ({\bf x
},{\bf y},t)+\Psi ({\bf x},t)\Psi ({\bf y},t)\right] \, , \label{psi}\\
i\hbar \frac{\partial }{\partial t}\Phi ({\bf x},{\bf y},t)&=& H_{
{\rm 2B}}({\bf x},{\bf y})\Phi ({\bf x},{\bf y},t)
+V({\bf x}-{\bf y})\Psi ({\bf x},t)\Psi ({\bf y},t) \, ,
\end{eqnarray}

\noindent where $H_{{\rm 2B}}({\bf x},{\bf y})=H_{{\rm 1B}}({\bf x})+H_{{\rm 1B}}(
{\bf y})+V({\bf x}-{\bf y})$. The first-order approach also includes
the appropriately truncated equations of motion for $\Gamma
({\bf x},{\bf y},t)$ and $\langle \psi ^{\dagger }({\bf
  x})\psi ({\bf y})\psi ({\bf z})\rangle^{c}$. We note \cite{TKKB02}, however,
that if $\Gamma ({\bf x},{\bf y},t)$ and $\langle \psi ^{\dagger }({\bf
  x})\psi ({\bf y})\psi ({\bf z})\rangle^{c}$ vanish initially,
they will not evolve. Thus, we neglect these two cumulants on the
right-hand side of Eq.(\ref{psi}).
Back substitution of the
formal solution of the linear equation of motion for $\Phi ({\bf
  x},{\bf y},t)$ leads to a closed equation of motion for the mean
field $\Psi ({\bf x},t)$:

\begin{equation}
i\hbar \frac{\partial }{\partial t}\Psi ({\bf x},t)=H_{{\rm 1B}}({\bf x}
)\Psi ({\bf x},t)+\Psi^{*} ({\bf x},t) \int_{t_{0}}^{t}dt_{1} (2\pi
\hbar)^3 \langle 0|T_{{\rm
      2B}}^{(+)}(t,t_{1})|0 \rangle \Psi^{2} ({\bf x},t_1) \, ,
\label{NMNLSE}
\end{equation}

\noindent where
\begin{equation}
T_{{\rm 2B}}^{(+)}(t,t_{1})=V\delta (t-t_{1})+VG_{{\rm 2B}}^{(+)}(t,t_{1})V
\end{equation}

\noindent is a retarded two-body time-dependent transition matrix. The
time-dependent transition matrix is expressed
in terms of the retarded two-body Green's function $G_{{\rm
    2B}}^{(+)}(t,t_{1})$, which is defined in a usual way through

\begin{equation}
\left( i\hbar \frac{\partial }{\partial t}-H_{{\rm 2B}}\right) G_{{\rm 2B}}^{(+)}(t,t_{1})=\delta (t-t_{1}) \, .
\label{G2B}
\end{equation}

\noindent The two-body physics enters the first-order cumulant approach through
the two-body transition matrix. Thus, the binary collisions are
described non-perturbatively. In Eq.(\ref{NMNLSE}), the transition
matrix is sandwiched between two plane waves of zero (relative) momentum (denoted
as $|0 \rangle $). This reflects the short range of the microscopic interatomic potential $V$ as
compared to the length scales relevant to the mean field $\Psi ({\bf
  x},t)$. In practical terms, Eq.(\ref{G2B}), describing the two-body
dynamics, can be solved by making use of a separable form of the
two-body interaction potential (for details, see
\cite{TKTGKB03,KGTKSAGETPSJ03}), which is a rather good approximation
for low-energy collisions. In the end, the two-body problem is
described in terms of a few collisional parameters (e.g. the
scattering length or the scattering resonance width), which can be
determined experimentally or theoretically.

Eq.(\ref{NMNLSE}) has the form of a non-Markovian non-linear Schr\"{o}dinger
equation. It can be shown that in the Markov approximation it reduces to
the well known time-dependent Gross-Pitaevskii equation \cite{TKKB02}. We note in passing the
limitations of the above formulation of the first-order cumulant
approach. Neglecting the initial one-body density matrix
of the non-condensate $\Gamma ({\bf x},{\bf y},t_0)$ implies that the
initial state must be described by a mean field. The above derivation
also assumes that the initial pair function $\Phi ({\bf x},{\bf
  y},t_0)$ vanishes, so no account of collisions before the initial
time is taken. Both assumptions are well satisfied in most present-day experiments.

Although $\Gamma ({\bf x},{\bf y},t)$ does not appear explicitly in
the equation of motion (\ref{NMNLSE}) (which reflects the assumption
that the non-condensed fraction does not affect the dynamics of the
mean field), the atoms lost from the condensate are transferred into
the non-condensate. Having solved for the evolution of the mean field and the
pair function, we can now use them to determine the evolution of the
non-condensed fraction by retaining only these terms in the full
equation of motion for $\Gamma ({\bf x},{\bf y},t)$
\cite{TKTGKB03}. The resulting dynamical equation has the following form:

\begin{equation}
i\hbar \frac{\partial }{\partial t} \Gamma ({\bf x},{\bf y},t)=\bigg\{
  H_\mathrm{1B} ({\bf x}) \Gamma ({\bf x},{\bf y},t) + \int \, d\mathbf{z}
  V({\bf x}-{\bf z})  \Phi^{*} ({\bf z},{\bf y},t) \left[  \Phi ({\bf
  x},{\bf z},t) + \Psi ({\bf x},t) \Psi ({\bf z},t) \right] \bigg\} -
  \bigg\{ {\bf x} \leftrightarrow  {\bf y} \bigg\}^{*} \, .
\label{gamma}
\end{equation}  

\noindent It can be shown that for an initially undepleted condensate,
i.e. $\Gamma ({\bf x},{\bf y},t_0)=0$, the solution of
Eq.(\ref{gamma}) can be expressed in terms of the pair function as

\begin{equation}
\Gamma ({\bf x},{\bf y},t)=\int d\mathbf{z} \, \Phi ({\bf x},{\bf z},t)
\Phi^{*} ({\bf y},{\bf z},t) \, .
\label{solution}
\end{equation}

\noindent Consequently, the non-condensed fraction stems from the
build-up of pair correlations. In turn, these correlations can appear
either in the scattering continuum (as correlated pairs of relatively
fast atoms) or in the bound part of the two-body spectrum, as
molecules \cite{TKTGKB03}. Given Eq.(\ref{solution}), the
total number of atoms $N$ can be shown to be a constant of motion:

\begin{equation}
N=\int d\mathbf{x} \left[ \Gamma ({\bf x},{\bf x},t) + |\Psi ({\bf
    x},t)|^2 \right] \, .
\end{equation}

\begin{figure}[htb]
\begin{center}
\includegraphics[width=8cm,clip]{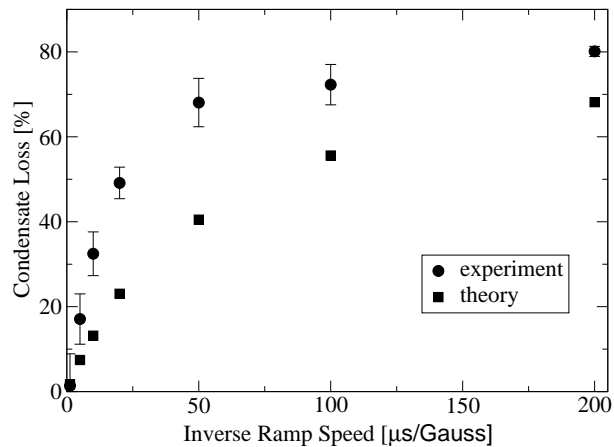}
\caption{
\label{loss}
Loss of atoms from a $^{85}$Rb condensate at the end of a linear 
magnetic field ramp across the 155 Gauss Feshbach resonance as a function 
of the inverse ramp speed. The circles with error bars are the experimental 
data from Ref.~\protect\cite{Cornish00}. The squares are determined 
from Eq.~(\ref{NMNLSE}) using a cylindrical trap potential with
the experimental trap frequencies.}
\end{center}
\end{figure}

As an application of the cumulant method
we study the experiment of Ref. \cite{Cornish00} performed with a
$^{85}$Rb condensate at JILA. In this experiment, the magnetic field
was swept across a Feshbach resonance, starting from the side of the positive scattering length. A loss
of atoms from the condensate was observed, which depended on the
ramp speed of the magnetic field strength. In applying the cumulant approach to
this case, special care has to be taken to describe properly the
underlying two-body dynamics in the vicinity of a Feshbach resonance \cite{TKKG03}. We have
solved the non-Markovian non-linear Schr\"{o}dinger equation
(\ref{NMNLSE}) to simulate the actual experimental procedure. At the
end of the ramp we record the decrease in the norm of the mean field $\Psi
({\bf x},t)$ and compare it with the experimentally measured condensate
loss. The results are shown in Figure \ref{loss}.

We have employed the non-Markovian equation of the first-order
cumulant approach to study other situations where significant loss of
condensate atoms due to a dynamic perturbation is observed. The method
has been successful in predicting the rate of atom loss in
four-wave mixing experiments \cite{TKKB02}. We have also analysed \cite{TKTGKB03}
the experiment pioneering the use of magnetic-field pulses for the
association of molecules, where coherent atom-molecule oscillations
have been observed \cite{Donley02}. In this context, the physical
consequences of an
extremely long-range nature of molecules created have been pointed out
\cite{TKTGPSJKB03} and the process of adiabatic association of dimer
molecules through
Feshbach resonance crossings has been analysed in detail
\cite{KGTKSAGETPSJ03}. The second-order approach has been used to
determine the mean field energy associated with three-body collisions
in Bose condensates \cite{TK02}.

There are other situations of experimental and theoretical interest
where the mean-field dynamical description can be expected to fail due to
the significance of pair correlations. In the context of optical
lattices, the cumulant approach could be a useful tool in describing
the dynamics of the superfluid bosonic gas in the vicinity of the
quantum phase transition, where quantum depletion starts to grow. An
area relatively unexplored experimentally is the dynamic manipulation
of magnetically-tunable interactions, not necessarily in the close vicinity of
a Feshbach resonance \cite{Abdullaev03}. The question of
adiabaticity of the evolution with respect to particle loss from the
condensate is an important one in this context. Finally, one can study
situations where a Bose condensate is
subjected to rapid variations of an external (e.g. trapping)
potential. One of the interesting proposals in this field is a
delta-kicked oscillator setup, employed in the studies of quantum
chaos \cite{Gardiner00}.

From the viewpoint of the classical field methods, the cumulant
approach also may offer an interesting perspective. The non-Markovian
equation (\ref{NMNLSE}) can
be viewed as a consistent way of including quantum corrections in the
classical-field treatment. As such, it can be used to test the validity
of replacing quantum field operators by complex-valued functions,
which is the backbone of the latter approach \cite{Svistunov91,Damle96,Goral01,Davis01,Sinatra01,Goral02,Schmidt03}.

To conclude: in this paper we have reviewed a new method of tackling
the dynamics of correlations in a Bose condensate, in terms of
noncommutative cumulants. This framework delivers an accurate but
computationally feasible approach to the beyond-mean-field
evolution in a Bose-condensed gas. The method has been used to
address several issues of current interest.
Finally, we have pointed out several new applications and a
possible link to the classical-field methods for partly condensed Bose gases.

We are particularly grateful to Eleanor Hodby, Simon Cornish and Carl Wieman 
for providing the experimental data in Fig.~\ref{loss}. We would also like to thank Simon
Gardiner and Paul Julienne for many interesting discussions. This
research has been supported by the European Community Marie Curie
Fellowship under Contract no.\ HPMF-CT-2002-02000 (K.G.), a University
Research Fellowship of the Royal Society (T.K.), 
Deutsche Forschungsgemeinschaft (T.G.). K.B. thanks the Royal Society
and the Wolfson Foundation.


\begin{thebibliography}{99}
\bibitem{note}
  We describe the state of the many-body system with the help of a
  statistical operator $\rho(t)$ and the expectation value of an
  arbitrary operator $A$ in this state is given by $\langle A
  \rangle_t=\mbox{Tr}[\rho(t)A]$. The evolution of such an expectation value
  is then determined by the Schr\"odinger equation $i\hbar
  \frac{\partial }{\partial t} \langle A\rangle_t=\langle [A,H]\rangle_t$.
\end{thebibliography}
\end{document}